\documentclass[10pt,journal,comsoc]{IEEEtran}
\usepackage{url}
\usepackage{caption}
\usepackage[noadjust]{cite}
\usepackage{graphicx}
\usepackage{multirow}
\usepackage{caption}
\usepackage{subcaption}
\usepackage{comment}
\usepackage{xspace}
\usepackage{caption}
\usepackage{subcaption}
\usepackage{makecell}
\usepackage{ragged2e}
\usepackage{rotating}
\usepackage{lscape}
\usepackage[table]{xcolor} 
\usepackage{nomencl}
\usepackage{mwe}
\usepackage{makecell}
\usepackage{balance}
\usepackage[utf8]{inputenc}
\usepackage{ragged2e}
\usepackage{booktabs}% http://ctan.org/pkg/booktabs

\usepackage{pifont}
\usepackage{tabulary}
\usepackage{comment}
\usepackage{lipsum}
\usepackage{array, tabularx, multirow, booktabs, diagbox}
\usepackage{adjustbox}

\usepackage{amsmath,amssymb,amsfonts}
\usepackage{algorithmic}
\usepackage{graphicx}
\usepackage{textcomp}
\def\BibTeX{{\rm B\kern-.05em{\sc i\kern-.025em b}\kern-.08em
    T\kern-.1667em\lower.7ex\hbox{E}\kern-.125emX}}

\begin{document}
%\title{Machine Learning for Wireless Metaverse: Overview, Key Requirements, and Challenges}
\title{Machine Learning for Wireless Metaverse: Fundamentals, Use Case, and Future Directions}

\author{Latif~U.~Khan,~\IEEEmembership{Member,~IEEE},~Ibrar~Yaqoob,~\IEEEmembership{Senior~Member,~IEEE},~Khaled~Salah,~\IEEEmembership{Senior~Member,~IEEE},~Choong~Seon~Hong,~\IEEEmembership{Senior~Member,~IEEE},~Dusit~Niyato,~\IEEEmembership{Fellow,~IEEE},~Zhu~Han,~\IEEEmembership{Fellow,~IEEE},~Mohsen~Guizani,~\IEEEmembership{Fellow,~IEEE}

\IEEEcompsocitemizethanks{
\IEEEcompsocthanksitem L.~U.~Khan~is with the Abu Dhabi University, United Arab Emirates and also with the Department of Machine Learning, Mohamed Bin Zayed University of Artificial Intelligence, United Arab Emirates.
\IEEEcompsocthanksitem M.~Guizani is with the Department of Machine Learning, Mohamed Bin Zayed University of Artificial Intelligence, United Arab Emirates.
\IEEEcompsocthanksitem I. Yaqoob is with the Artificial Intelligence and Cyber Futures Institute, Charles Sturt University,  Bathurst, NSW 2795, Australia.
\IEEEcompsocthanksitem K. Salah is with the Department of Computer and Information Engineering, Khalifa University, United Arab Emirates.
\IEEEcompsocthanksitem C.~S.~Hong is with the Department of Computer Science and Engineering, Kyung Hee University, South Korea.
\IEEEcompsocthanksitem D. Niyato is with the School of Computer Science and Engineering, Nanyang Technological University, Singapore.
\IEEEcompsocthanksitem Zhu Han is with the Electrical and Computer Engineering Department, University of Houston, Houston, TX 77004 USA.

\IEEEcompsocthanksitem Corresponding Author: Latif U. Khan \{latif.u.khan2@gmail.com\}

%\IEEEcompsocthanksitem Walid Saad is with the  Wireless@VT, Bradley Department of Electrical and Computer Engineering, Virginia Tech, Blacksburg, VA 24061 USA.
%, and also with the Department of Computer Science and Engineering, Kyung Hee University, Seoul 02447, South Korea. 

}}

\markboth{IEEE Internet of Things Magazine}{}%

\maketitle

% \maketitle

% }

% \thanks{
% 	}
% }

 %The paper headers
% \markboth{IEEE Communications Magazine}{}%

% Traditional machine learning is based on the migration of data from massively distributed IoT devices to a centralized cloud for training and thus, poses serious security and privacy concerns.

% \IEEEcompsoctitleabstractindextext{%
% \justify
\begin{abstract} 
Today's wireless systems are posing key challenges in terms of quality of service and quality of physical experience. Metaverse has the potential to reshape, transform, and add innovations to the existing wireless systems. A metaverse is a collective virtual open space that can enable wireless systems using digital twins, digital avatars, and interactive experience technologies. Machine learning (ML) is indispensable for modeling twins, avatars, and deploying interactive experience technologies. In this paper, we present the role of ML in enabling metaverse-based wireless systems. We discuss key fundamental concepts for advancing ML in the metaverse-based wireless systems. Moreover, we present a case study of deep reinforcement learning for metaverse sensing. Finally, we discuss the future directions along with potential solutions.

\end{abstract}


%In hierarchical federated learning, initially sub-global models are computed at small cell base stations in an iterative manner similar to traditional federated learning. Then, the sub-global models are sent to the macro base station where global model aggregation takes place. Finally, global model updates are sent back to the small cell base stations which further disseminate them to end-devices. The main advantage of hierarchical FL is reuse of the already occupied frequency bands by other users within small cells.

\section{Introduction}
\begin{figure*}[!t]
	\centering
	\captionsetup{justification=centering}
	\includegraphics[width=18cm, height=18cm]{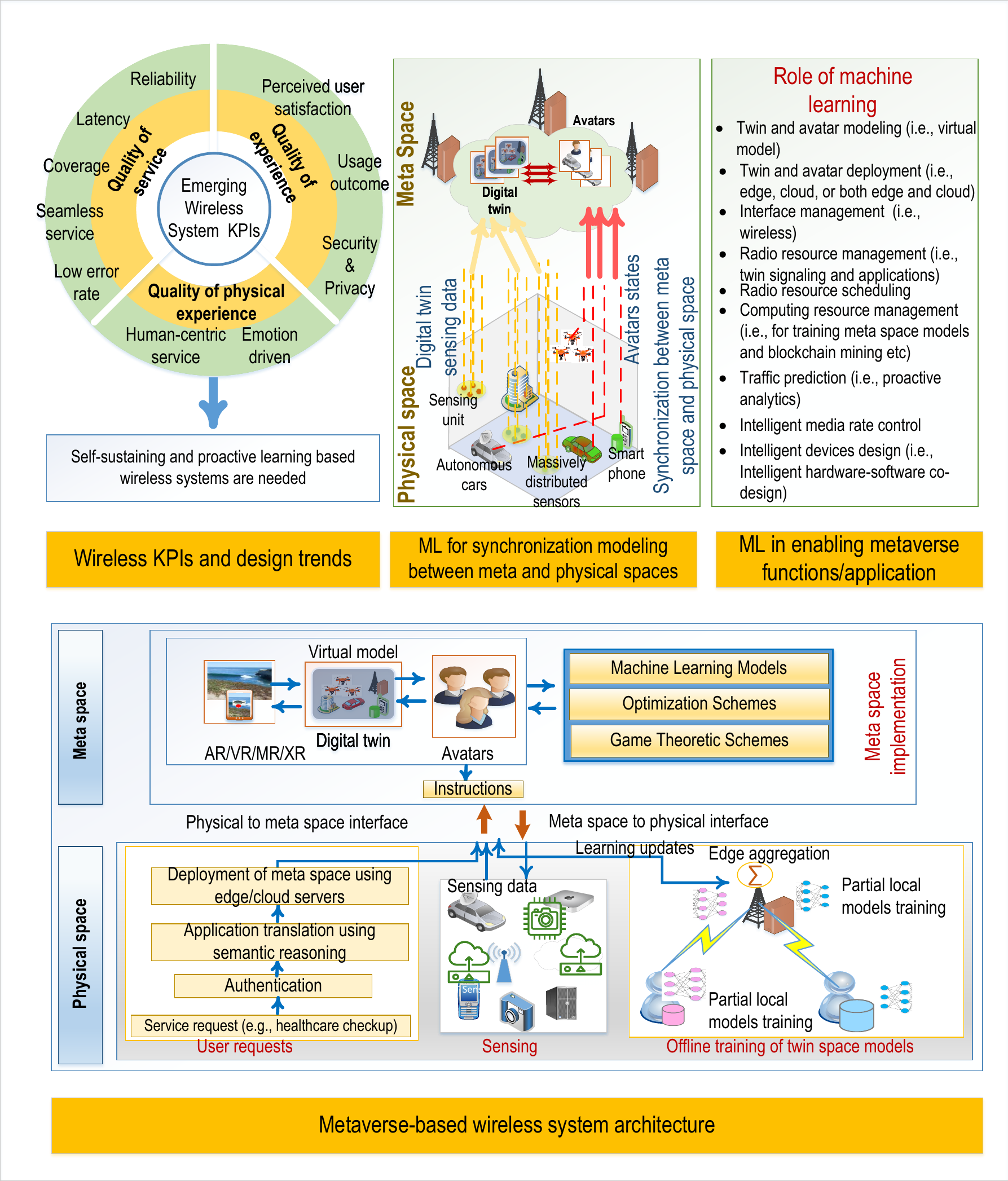}
	\caption{\textcolor{black}{Wireless/IoT systems KPIs, metaverse architecture, and role of machine learning.}}
	\label{fig:architecture}
\end{figure*}

\setlength{\parindent}{0.7cm}\textcolor{black}{Emerging Internet of Thing (IoT) applications, such as digital healthcare, intelligent transportation systems, Industry 4.0, and smart homes, are characterized by a wide variety of diverse user-defined metrics (e.g., quality of physical experience) and traditional metrics (e.g., latency and reliability), as shown in Fig.~\ref{fig:architecture}. The existing wireless/IoT systems fall short of meeting these diverse requirements \cite{khan2022digital,8869705}. A metaverse can be used to enable a wireless system to meet a variety of requirements \cite{khan2022metaverse}. The metaverse has the potential to effectively enable proactive online-learning and self-sustainability for wireless systems. Self-sustainability will help wireless systems operate with the least possible intervention from operators/users, whereas proactive learning will help in efficient optimization of wireless system resources for various applications (e.g., healthcare systems for lung diseases)/functions (e.g., wireless resource allocation). Proactive learning is necessary because many wireless system applications (e.g., digital healthcare and intelligent transportation system) have strict latency constraints. A typical IoT system has a wide variety of players, such as edge/cloud computing resources, wireless resources, device computing resources, and core network resources. Additionally, there will be a massive number of devices that made the system more complex for resource management. There is a need for seamless interaction among these players to enable various applications. Therefore, proactive learning prior to a user request can enable us to train machine learning (ML) models that can be used for future user requests by efficiently optimizing the application resources.} \par

\textcolor{black}{There are two main aspects, \emph{wireless for metaverse} and \emph{metaverse for wireless} \cite{khan2022metaverse,lin20236g}. The use of wireless technologies to enable the metaverse is referred to as wireless for the metaverse, whereas metaverse for wireless is the use of the metaverse to enable wireless applications. A meta space represents a virtual model and comprises of digital twins (e.g., hospitals and base stations) and digital avatars (e.g., mobile users and moving autonomous cars)\footnote{For more information about the role of avatars in metaverse, please see \cite{khan2022metaverse}.}. A typical metaverse-based wireless system consists of two spaces (i.e., shown in Fig.~\ref{fig:architecture}): meta space and physical interaction space \cite{khan2022metaverse}. The physical interaction space consists of all physical devices, edge/cloud servers, and other network switches, and is responsible for actual communication and computation. On the other hand, the logical space (i.e., meta space) consists of digital twins and digital avatars.} \textcolor{black}{Additionally, meta space will consider interactive experience technologies (i.e., virtual reality (VR), augmented reality (AR), mixed reality (MR), and extended reality (XR)), mathematical optimization, game theory, and machine learning (ML) for performance enhancement. ML is can be widely used compared to mathematical model (i.e., convex optimization schemes) because of their feature to model both convex and non-convex problems. Additionally, ML provides more flexibility in modeling of dynamic systems and generation of data/scenarios for new applications that are not yet deployed using generative ML.} \par
\textcolor{black}{The role of a digital avatar in the metaverse is to address the wireless system uncertainties due to mobile devices/users \cite{khan2022metaverse,ynag2022fusing,wang2022survey}. A typical digital twin can represent static entities (e.g., smart homes and base stations) of the wireless system. However, wireless systems have mobile users and devices that will significantly affect the performance. Therefore, there is a need to tackle this issue. \textcolor{black}{To account for statistical uncertainties in the wireless systems due to mobile devices, we need digital avatars. There is a need for effective modeling techniques to create avatars and digital twins.} One can use mathematical modeling; however, it suffers from the limitation of inaccurate results. We can also use experimental modeling to model avatars and twins. However, it also has limitations of experimental equipment and human errors. To address these limitations, one can use ML for modeling (i.e., data-driven modeling) of meta space entities. Various functions that can be performed using ML are presented in Fig.~\ref{fig:architecture}.} Various works in literature considered metaverse~\cite{ning2021survey,huynh2022artificial, ynag2022fusing,khan2022metaverse}. The work conducted in \cite{ning2021survey} presents the concept, recent advances, and open challenges. Huynh-The \textit{et al.} in \cite{huynh2022artificial} discussed artificial intelligence (AI) to enable the metaverse. Specifically, the authors presented an overview of various AI techniques as well as other key enablers (e.g., computer vision and natural language processing) of the metaverse. Another work \cite{ynag2022fusing} discussed the fusion of AI and blockchain towards realizing a metaverse. \textcolor{black}{The work in \cite{khan2022metaverse} presented the vision as well as the architecture of using wireless systems by metaverse. In contrast to the studies conducted in \cite{ning2021survey, huynh2022artificial, ynag2022fusing,khan2022metaverse}, our work focuses on the key role, a use case, and challenges in advancing ML for the metaverse-based wireless networks.} Our key contributions are as follows.\par

\begin{figure*}[!t]
	\centering
	\captionsetup{justification=centering}
	\includegraphics[width=17cm, height=14cm]{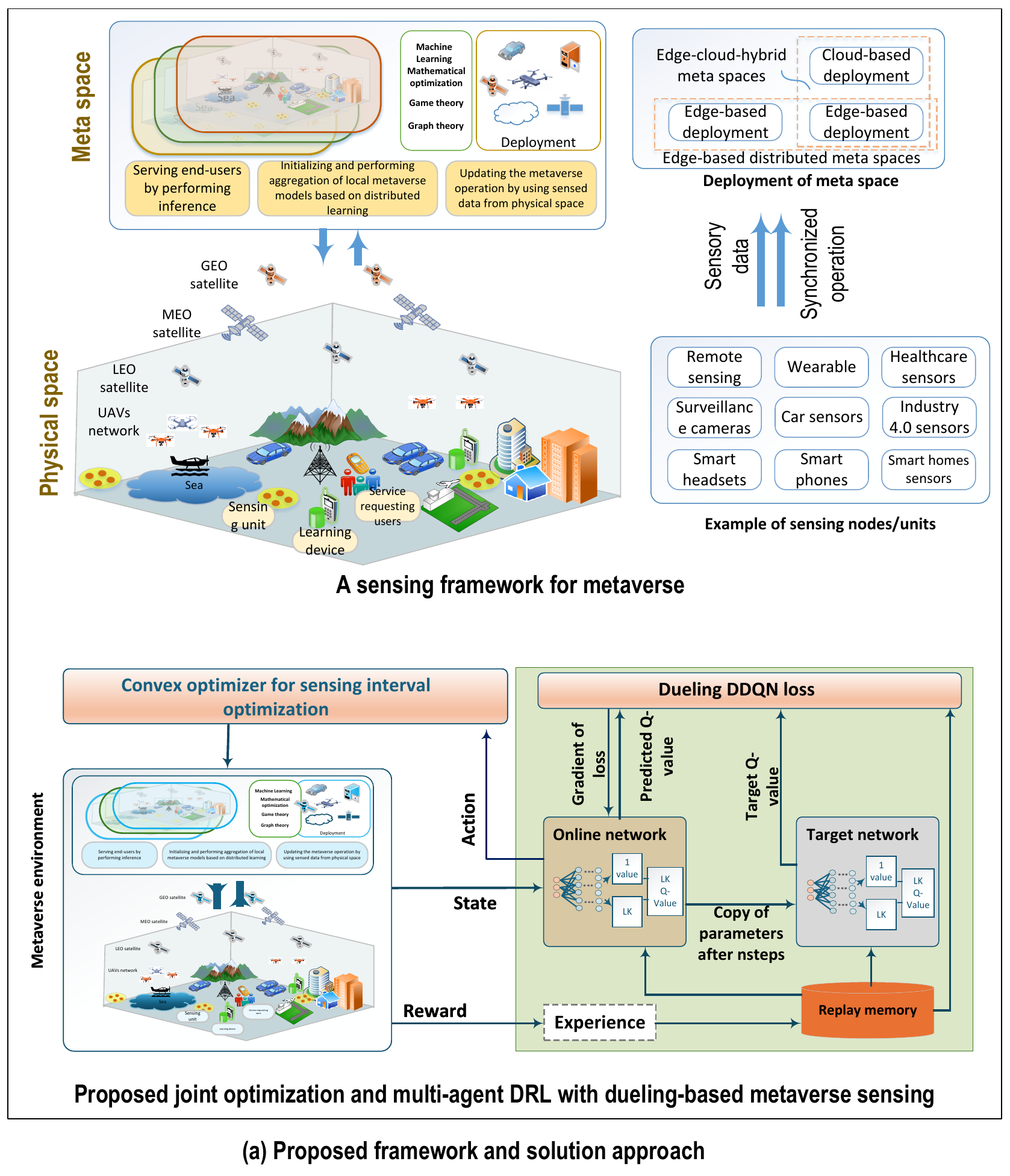}
	\caption{Multi-agent DDQN with dueling and optimization for sensing in metaverse.}
	\label{fig:case_A}
\end{figure*}

\begin{itemize}
    \item We present an overview of the metaverse and its architecture. The high-level architecture consists of meta space and physical space, as well as interfaces for communication. We discuss the role of ML in enabling metaverse applications and functions (e.g., avatar modeling and air interface modeling). 
    \item We identify key fundamental aspects of metaverse design and the role of ML in enabling metaverse. Two design aspects (i.e., ML for enabling metaverse and metaverse ML models for end-use applications) of metaverse design are presented. Additionally, we present a case study double deep Q-network with dueling as well as optimization for metaverse sensing. 
    \item We present future directions that exist in advancing ML for the metaverse.
\end{itemize}

\begin{figure*}[t!]
	\centering
		\begin{subfigure}[b]{0.3\textwidth}
		\centering
		\includegraphics[width=2.4in,height=1.4in]{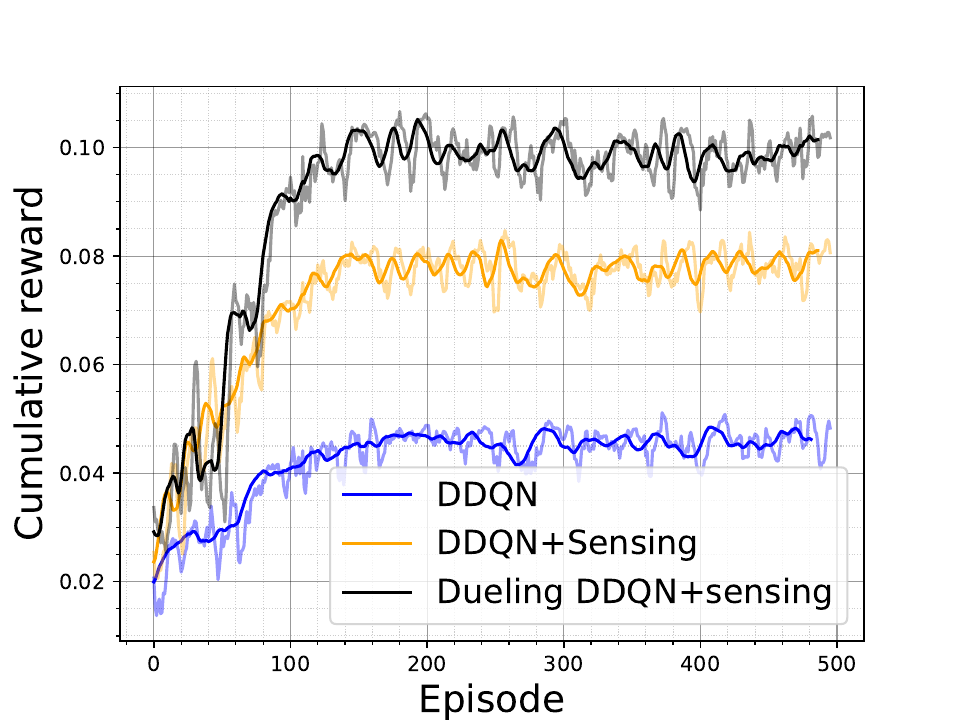}
		\caption{}	
	\end{subfigure}
	\hfill
	\begin{subfigure}[b]{0.3\textwidth}
		\centering
		\includegraphics[width=2.4in,height=1.4in]{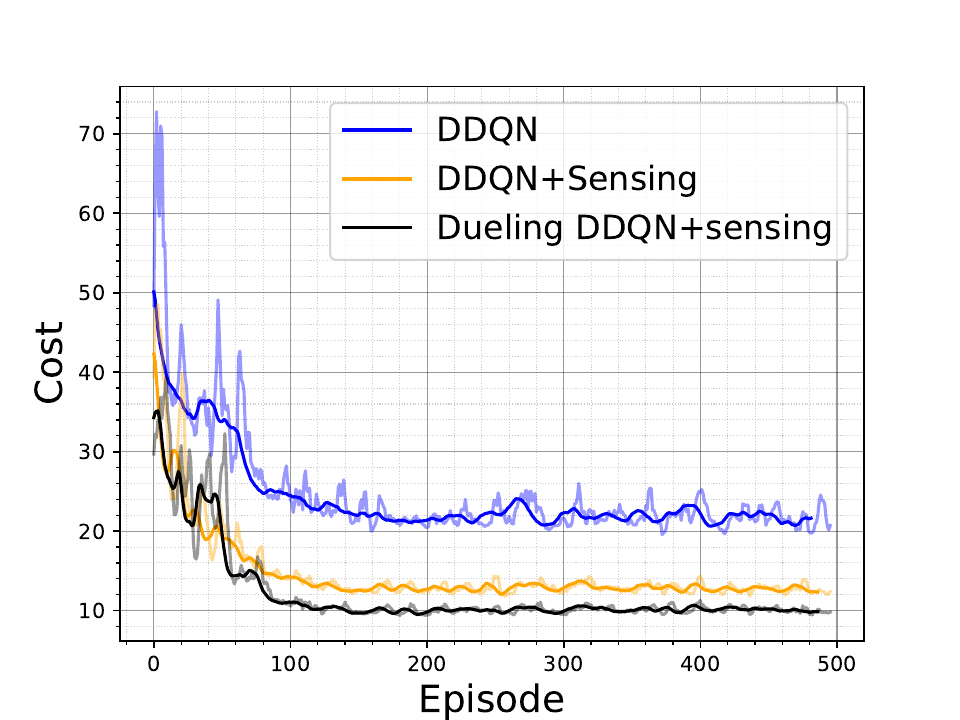}
		\caption{}	
	\end{subfigure}
	\hfill
	\begin{subfigure}[b]{0.3\textwidth}
		\centering
		\includegraphics[width=2.4in,height=1.4in]{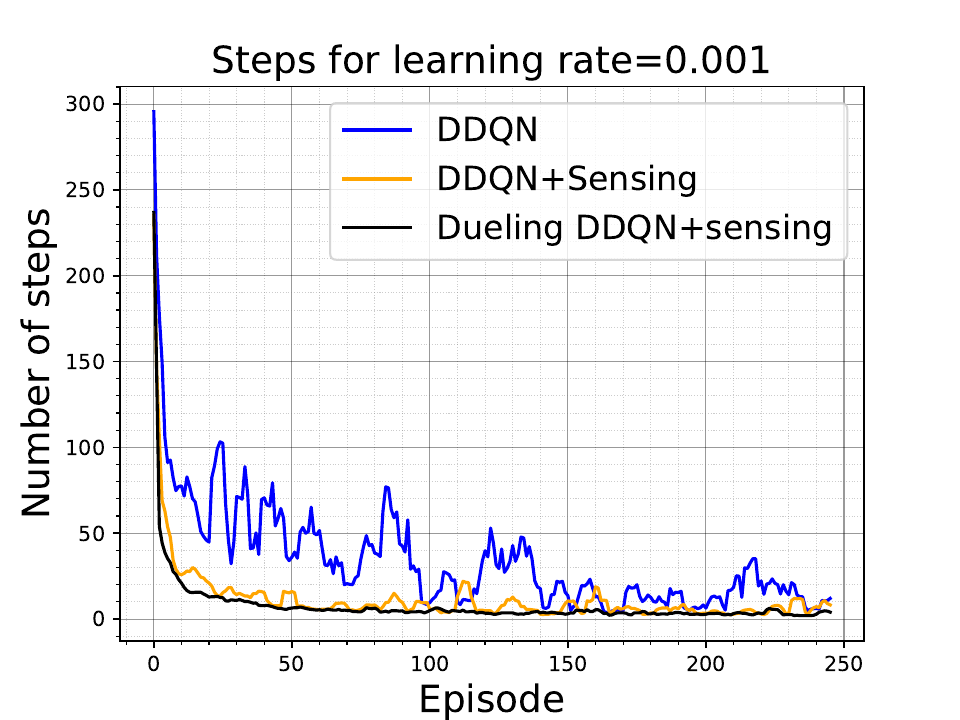}
		\caption{}
	\end{subfigure}
 
	\centering
	\caption{(a) Cumulative reward vs. episodes for various schemes, (b) cost vs. episodes for various schemes, and (c) number of steps needed to reach QoS for all sensing units using various schemes.}
	\label{fig:simlation}
\end{figure*}

\section{Fundamentals and Use Case of ML for Metaverse}

\subsection{Fundamentals}
A typical metaverse-based wireless system has two main parts: (a) meta space (i.e., virtual model of the physical world) and (b) physical space (i.e., actual world). These spaces interact with each other for better resource management to serve end-users of various applications. In a metaverse-based wireless system, there are three main tasks: (a) learning, (b) sensing, and (c) service requests \cite{khan2022metaverse}. In the meta space, various meta space models are needed to serve the end-users instantly. \textcolor{black}{To do so, one can use federated learning (FL) due to its better privacy-preserving nature compared to centralized ML. Moreover, FL is more suitable for frequent data generation scenarios. FL is based on training end-devices models and then sharing them with the meta space which combines them to generate a global model. Later, this global model is shared back with the physical space entities. This process takes place iteratively. Sharing of a learning model instead of whole data has less overhead and thus FL is more suitable for frequent data-generating scenarios compared to centralized ML that is based on sending of entire data. On the other hand, there are many sensors in the physical world that sense and then share the sensory data with the meta space. There can be many ways to deploy the sensors. One way is to deploy a massive number of sensors. Note here that the density of sensors will determine the accuracy and effectiveness. For communication, one can use orthogonal frequency division multiple access or time division multiple access or hybrid model using both the later schemes. A low density will have less communication overhead but at the cost less sensing accuracy.} \textcolor{black}{Conversely, a high density of sensors will have better accuracy, but at the cost of high communication overhead. Therefore, one must make a tradeoff between sensing accuracy, synchronization effectiveness, and communication overhead.}\par 
For effective sensing in metaverse, one can use the concept of massive distribution of sensing devices. Modeling massive distribution is very challenging. Therefore, one should consider them continuous similar to the work in \cite{hashash2022towards}. Then, the sensing units will be divided into sensing units. The data generated by every sensing unit will be then computed to yield the latency and other parameters used for performance evaluation parameters (e.g., transmission energy) used in communication. Other than sensing and learning, there is a need to model service requests. To simultaneously model service requests, learning, and sensing, one can use the concept of puncturing \cite{khan2024tcs,chen2022coexistence}. We can divide the resource blocks into mini resource blocks. Initially, we can assign multiple mini resource blocks to learning and sensing units/devices. The main reason for this assignment is the frequent nature of learning and sensing compared to service requests. Compared to learning and sensing, service requests are less frequent, they need more reliable and latency-aware communication. Due to its less frequent nature, it would not be a desirable solution to allocate dedicated resources during time slots to service-requesting devices. Therefore, one can use the concept of puncturing that makes the transmit power of sensing and learning devices zero in the punctured slots used by the requesting devices. \textcolor{black}{After receiving the sensed data by the meta spaces deployed at the network edge, updating of the states of meta spaces takes place. Next, we present a novel case study of sensing in metaverse.}

\subsection{Use Case: Multi-Agent Deep Reinforcement Learning with Dueling and Optimization for Sensing in Metaverse}
\textcolor{black}{We present a case study for sensing in metaverse as shown in Fig.~\ref{fig:case_A}. There are two main spaces: (a) physical space where sensing takes place and (b) meta space that performs overall management of the system. The sensed data from the physical space should be timely shared with the meta space for up-to-date operation\footnote{For more information about sensing in the architecture of wireless metaverse, please refer to \cite{khan2022metaverse}.}. In the physical space, a massive number of sensors are deployed. To model communication of these massively deployed sensors, one can have two ways. The first one is to model for a massive number of sensors that is very challenging. This approach will lead to a significantly higher computational complexity. To avoid the complexity, we assume a continuous distribution similar to the work in \cite{hashash2022towards} for effective sensing. The next step is to divide the whole region into discrete regions and then model communication for them. To address this, one can divide the massively and continuously distributed sensors into discrete regions, and then model their communication with the meta space \cite{hashash2022towards}. This approach will have a less complexity compared to modeling a massive number of sensors individually. A set $\mathcal{N}$ of $N$ sensing units are considered for sensing in the metaverse. For every sensing region, we consider a communication overhead, which is the function of sensing interval, infinitesimal volume, and unit volumetric density. The data generated by $n^{th}$ sensing unit can be given by $\mathcal{D}_{n}=\chi_{g}\epsilon\eta(x,y,z),~\forall{n \in \mathcal{N}}$, where $\chi_{n}$ and $\epsilon$ are the sensing interval and the infinitesimal volume. $\eta(x,y,z)$ is the unit volumetric density \cite{toumpis2006optimal}.} It is clear that sensing intervals should be chosen wisely, i.e., very large sensing intervals result in more sensing overhead and thus high latency and vice versa. Meanwhile, the sensing interval should not be very small which will lead to less synchronization between the physical space and the meta space. Based on the aforementioned discussion, it is necessary that we should make a tradeoff between the overhead and synchronization for sensing in the metaverse. On the other hand, the sensing delay must be less than a certain maximum threshold to ensure the up-to-date operation of the wireless metaverse. We consider transmission latency of sensing as a cost function (i.e., $\mathcal{C}$). Meanwhile, consider multiple meta spaces to serve physical space. The reasons for using distributed meta spaces are robustness, low latency, and more context-awareness, among others. Furthermore, we consider the following constraints:
\begin{itemize}
    \item \textcolor{black}{For every sensing unit $n$, there should be a maximum of one resource block $r$. Similarly, a resource block should be allocated to a maximum of one sensing unit.}
    \item \textcolor{black}{Every sensing unit should be associated with a single meta space. Moreover, every meta space should be assigned sensing units without exceeding their maximum serving capacity due to computing resources constraints.} 
    \item \textcolor{black}{For sensing in the metaverse, there should be a constraint of QoS in terms of maximum available transmission latency. Therefore, the transmission latency should be less than the maximum allowed threshold.}
    \item \textcolor{black}{The sensing intervals should be selected within the limits. Meanwhile, the summation of the sensing intervals should not exceed the maximum limit. The reason for using this maximum limit is the limitation of available computing resources for sensing in sensing units. }
\end{itemize}\par
\textcolor{black}{The formulated problem is to minimize the cost by optimization of the sensing interval, association of sensing units with the meta spaces, and the wireless resource allocation. Our formulated metaverse sensing problem is a mixed integer non-linear programming (MINLP) and non-convex problem. Therefore, one can not simply use convex optimization and combinatorial problem solutions (e.g., matching game-based solutions). To solve this problem, we propose a novel DDQN assisted by dueling and optimization-based solution as shown in Fig.~\ref{fig:case_A}. First, the process is defined as stochastic game and then Markov Decision Process (MDP) is used to model it due to the fact actions in partially under the control of agents. Later, we propose the use of DDQN for solving MDP. In our solution, the reward is taken equal to $1/\mathcal{C}$.} \textcolor{black}{The state of any agent deployed for sensing unit represents association and resource allocation. Note that dueling is very useful for scenarios with dynamic states \cite{zhao2019deep}. In dueling, a Q-value function is decomposed into two values: (a) state-value function and (b) advantage function. These values are then used to obtain a Q-value function. Other than dueling, we perform sensing interval optimization using a convex optimizer. This is due to the fact that the reward (i.e., $\frac{1}{\mathcal{C}}$) of the proposed scheme depends on sensing interval in addition to association and resource allocation. In addition to sensing interval optimization, the association and resource allocation are performed using dueling-based DDQN. Fig.~\ref{fig:simlation}a and Fig.~\ref{fig:simlation}b show the results for proposed schemes (i.e., DDQN+sensing+dueling and DDQN+sensing) and the traditional DDQN scheme. It is clear from Figs.~\ref{fig:simlation}a and ~~\ref{fig:simlation}b that the proposed schemes outperform the traditional DDQN. On the other hand, Fig.~\ref{fig:simlation}c shows the performance of the proposed scheme (i.e., DDQ+sensing+dueling) for achieving a QoS for various schemes. It is evident that our proposed scheme converges fast compared to other schemes. The reason for performance improvement of the proposed scheme compared to other baselines lies in the fact that proposed scheme optimizes sensing interval, association, and resource allocation. Optimizing sensing interval will results in less overall overhead and thus better performance. For all results, the performance trend is same, i.e., our proposed schemes (i.e., dueling+DDQN+sensing and DDQN+sensing) outperforms traditional DDQN. }

\begin{figure*}[!t]
	\centering
	\captionsetup{justification=centering}
	\includegraphics[width=16cm, height=6.5cm]{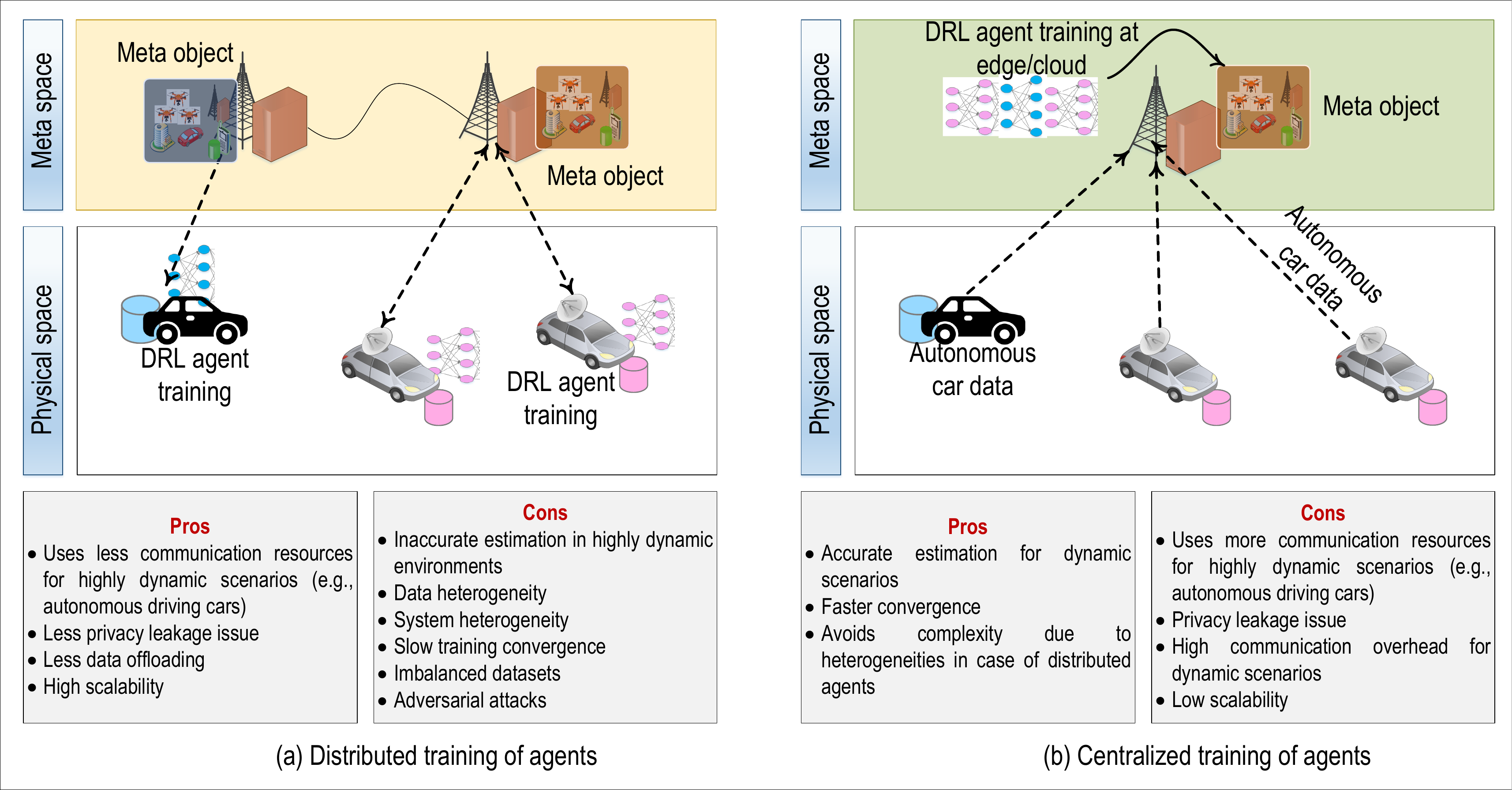}
	\caption{Meta object models training fashion: (a) distributed and (b) centralized. }
	\label{fig:training_fashion}
\end{figure*}

\begin{figure}[!t]
	\centering
	\captionsetup{justification=centering}
	\includegraphics[width=8cm, height=5cm]{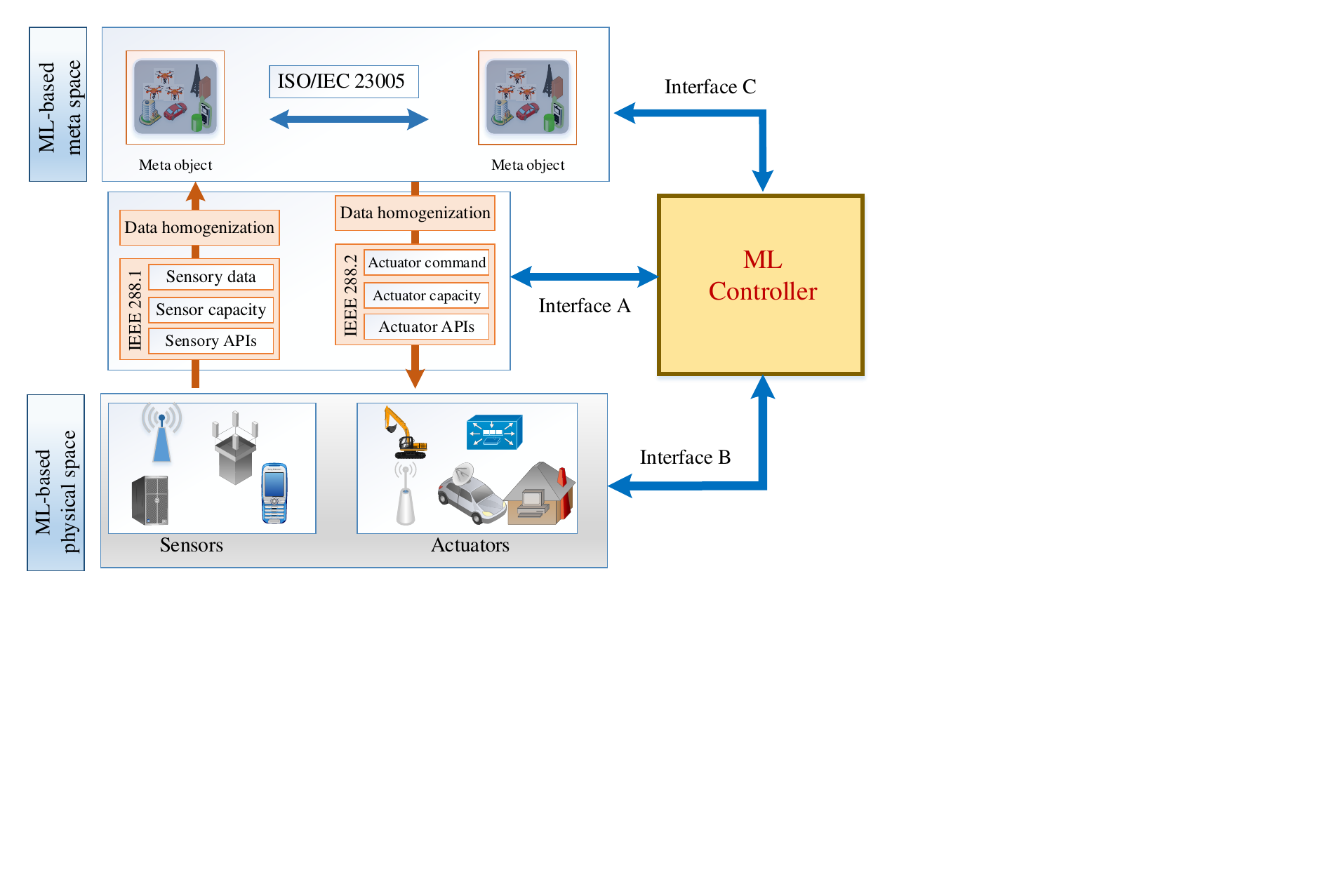}
	\caption{Standardization of ML-based metaverse}
	\label{fig:stadandization}
\end{figure}

\section{Conclusions and Future Directions}
\textcolor{black}{In this paper, we have discussed the role of ML in enabling metaverse-based wireless systems. We concluded that ML will play a key role in enabling metaverse-based wireless systems. Several key future directions, along with causes and possible solutions, that exist in advancing ML for metaverse-based wireless systems are presented as follows. }

\subsection{Gen-AI for Metaverse Content Generation}
{\em How do we model metaverse virtual entities?} Virtual modeling of the real-world metaverse entities for use in a meta space is challenging due to non-availability of data and their novel nature. One can mainly use Gen-AI to enable dynamic simulations, behavioral modeling, and system optimization. For dynamic simulations, one can use Gen-AI to generate simulations as well as data. For instance, if we want to simulate the real world, there is a need for sensory data (e.g., ITS images). Such data can be generated by generative adversarial networks. \textcolor{black}{Although generative adversarial networks can be used to generate data, there might be challenges, such as object diversity (e.g., pedestrians, different types of vehicles, and traffic signs), occlusion handling (e.g., pedestrians might be hidden by another vehicle), and high spatial resolution. Humans significantly affect the wireless communication, especially terahertz communication. For higher communication bands, humans cause significant attenuation, scattering, and signal absorption.} Therefore, one must carefully design and propose schemes to generate such kind of data. On the other hand, modeling of human impact on communication (e.g., Terahertz communication) is also challenging and needs careful design and attention. 

\subsection{Intelligent Resource Scheduling for Metaverse}
{\em How does one enable efficient resource scheduling for metaverse?} In a metaverse, there are a variety of entities. These entities are sensing nodes/units, requesting devices/users, and learning devices. All of these devices from the physical space communicate with the meta space. Modeling such an interaction is very challenging. One can have mini slots of resource blocks. Learning and sensing are more frequent compared to service requests. Therefore, one can use the concept of puncturing and risk-aware formulation to reflect the degradation effect due to puncturing on the cost function. In this approach, we can assign mini resource blocks to learning and sensing units. Upon service request (e.g., rending task for AR device) from users, one can do puncturing by making the transmit power zero in some of the mini resource blocks assigned to learning and sensing devices. These punctured resource blocks are assigned to service-requesting devices. \textcolor{black}{To perform such a kind of resource scheduling, one can use a deep reinforcement learning-based scheme (e.g., double deep Q-learning). Although DDQN performs well, there could be many novel schemes (e.g., dueling-based DDQN similar to case study A in Section II.1B) to improve the performance further. On the other hand, one can combine attention mechanisms with reinforcement learning to improve performance. In the case of attention, each agent focuses on the most relevant parts of the input space and is given a differentiated fit reward to improve the performance \cite{wang2022performance}.} Furthermore, to deploy learning agents, one can have several fashions, shown in Fig.~\ref{fig:training_fashion}. Various deployment modes offer different benefits and costs in terms of management as well as communication overhead. Therefore, one should make a trade-off between performance and cost.

\subsection{Standardization}
{\em How does one propose an efficient standard for ML-enabled metaverse for wireless systems?} Existing standards (e.g., ISO/IEC 23005 and IEEE 2888) of the metaverse mainly focus on interfaces for seamless connectivity between the real world and the virtual world \cite{wang2022survey}. ISO/IEC 23005 standards enable various metaverse business services. In these services, the virtual objects (e.g., virtual items and avatars) characteristics, association of rendered sensory effects, and audiovisual information leverages interactions between real worlds and virtual worlds. Specifically, ISO/IEC 23005 standards focus on sensory effects. On the other hand, IEEE 2888 standards enable the foundations of metaverse systems. IEEE 2888.1 and IEEE 2888.2 standards are used for exchange of actuator-related information and sensory information between the physical and virtual worlds, respectively. The ISO/IEC 23005 standards lack general-purpose interfaces for communication between the virtual and physical worlds to enable various emerging applications. Furthermore, while both the ISO/IEC 23005 and IEEE 2888 standards can provide numerous benefits, they do not explicitly address the role of machine learning in enabling the metaverse. \par
One can use ML to effectively enable metaverse-based wireless systems. For instance, modeling of meta objects for a particular application/function can be performed using supervised learning-based schemes, reinforcement learning-based schemes, and transfer learning-based schemes. For power control in an access network, one can use Q-learning. Therefore, it is clear that ML is inevitable to enable metaverse-based wireless systems. However, the existing architecture proposed by many metaverse works may not provide enough flexibility to support ML effectively. Therefore, there is a need to propose novel standardization schemes for the practical implementation of ML-based metaverse, as shown in Fig.~\ref{fig:stadandization}. Fig.~\ref{fig:stadandization} shows the general overview of using ML for enabling the metaverse. There is a need for an ML controller that will use ML to control the activities in the physical space, meta space, and interfaces. Note that existing standards should be used in addition to existing standards, such as IEEE 288.1, IEEE 288.2, and ISO/IEC 23005. There is a need to propose a novel standard for data homogenization in a metaverse. The need for data homogeneity arises due to the fact that different data-generating sources in the metaverse will have different forms that need to be transformed into a general form acceptable by the meta space \cite{khan2022metaverse}. This approach will enable a general design of meta space for various applications and will also enable a less time-consuming solution. Additionally, there is a need to standardize the interfaces, such as interface A, interface B, and interface C, as shown in Fig.~\ref{fig:stadandization}. The interfaces (i.e., interfaces A and B) between the ML controller and physical space must be real-time, whereas the interface between the ML controller and meta space can be real-time or non-real-time. For offline training of meta models in meta space, the interface C can be non-real-time, whereas the management of meta objects can be controlled by a real-time interface C. Due to the important role of ML in enabling the metaverse (as shown in Fig.~\ref{fig:stadandization}), there is a need to propose novel standards.

%\section{Conclusions}

%In this article, we have discussed the role of ML in enabling metaverse-based wireless systems. The role and a set of key requirements for ML towards enabling metaverse-based wireless systems are presented. We outlined and discussed in detail the key challenges that exist toward advancing ML for metaverse. Furthermore, causes and possible solutions to these challenges are presented. We conclude that ML will play a key role in enabling metaverse-based wireless systems.     

\bibliographystyle{IEEEtran}
\bibliography{Database}

\end{document}